\documentclass[prl,reprint,superscriptaddress,nofootinbib,twocolumn]{revtex4-1}

\usepackage[linktoc=page,colorlinks,urlcolor=blue,citecolor=blue,linkcolor=blue]{hyperref}
\usepackage{amssymb,amsmath}
\usepackage{graphicx}
\usepackage{natbib}
\usepackage{mathrsfs}
\usepackage{overpic}
\usepackage{gensymb}

\usepackage{pdfpages}
\makeatletter
\AtBeginDocument{\let\LS@rot\@undefined}
\makeatother

\providecommand{\e}[1]{\ensuremath{\times 10^{#1}}}

\usepackage{color}
\usepackage{upgreek}
\usepackage[letterpaper,textwidth=7in,top=.75in,bottom=.75in]{geometry}

\newcommand{\snl}{Sandia National Laboratories, Albuquerque, New Mexico 87185, USA}
\newcommand{\cint}{Center for Integrated Nanotechnologies, Sandia National Laboratories, Albuquerque, New Mexico 87123, USA}
\begin{document}
\title{A physically unclonable function using NV diamond magnetometry and micromagnet arrays}
\date{\today}
\author{Pauli Kehayias}
\email{pmkehay@sandia.gov}
\affiliation{\snl}
\author{Ezra Bussmann}
\affiliation{\snl}
\affiliation{\cint}
\author{Tzu-Ming Lu}
\affiliation{\snl}
\affiliation{\cint}
\author{Andrew M. Mounce}
\affiliation{\snl}
\affiliation{\cint}

\begin{abstract}
A physically unclonable function (PUF) is an embedded hardware security measure that provides protection against counterfeiting. Here we present our work on using an array of randomly-magnetized micron-sized ferromagnetic bars (micromagnets) as a PUF. We employ a 4 $\upmu$m thick surface layer of nitrogen-vacancy (NV) centers in diamond to image the magnetic fields from each micromagnet in the array, after which we extract the magnetic polarity of each micromagnet  using image analysis techniques. After evaluating the randomness of the micromagnet array PUF and the sensitivity of the NV readout, we conclude by discussing the possible future enhancements for improved security and magnetic readout. 

\end{abstract}
\maketitle

\section{Introduction}

Physically unclonable functions (PUFs) are physical entities that are intrinsically built into an object, are difficult to duplicate, and serve as a unique identifier (fingerprint) for the object \cite{maesPUFBook, herder2014, katzenbeisser2012, lim2005, PUFtaxonomy, sandiaPUFsPrimer}. A PUF relies on uncontrollable randomness inherent to the manufacturing process to provide the identifier for each object.  Many PUF implementations exploit random variations in semiconductor device fabrication, while others use alternative sources of randomness (such as fibers in a banknote or etch pit lengths in a compact disc). The PUF security comes from the fact that copying an individual one is difficult. To clone a particular PUF instance, a counterfeiter would need to use a prohibitively sophisticated fabrication method or make many instances before randomly producing a similar one. In addition to being easy to manufacture but difficult to copy, a PUF should be easy to characterize, have a reproducible output when characterized, be random and unique, and ideally be low-cost and resilient to the environment. In this work we investigate a PUF based on the random magnetization directions in an array of fabricated micron-sized ferromagnetic bars (micromagnets) on a silicon wafer, together with optical magnetic readout using nitrogen-vacancy (NV) defect centers in diamond.

\begin{figure*}[ht]
\begin{center}
\begin{overpic}[width=0.95\textwidth]{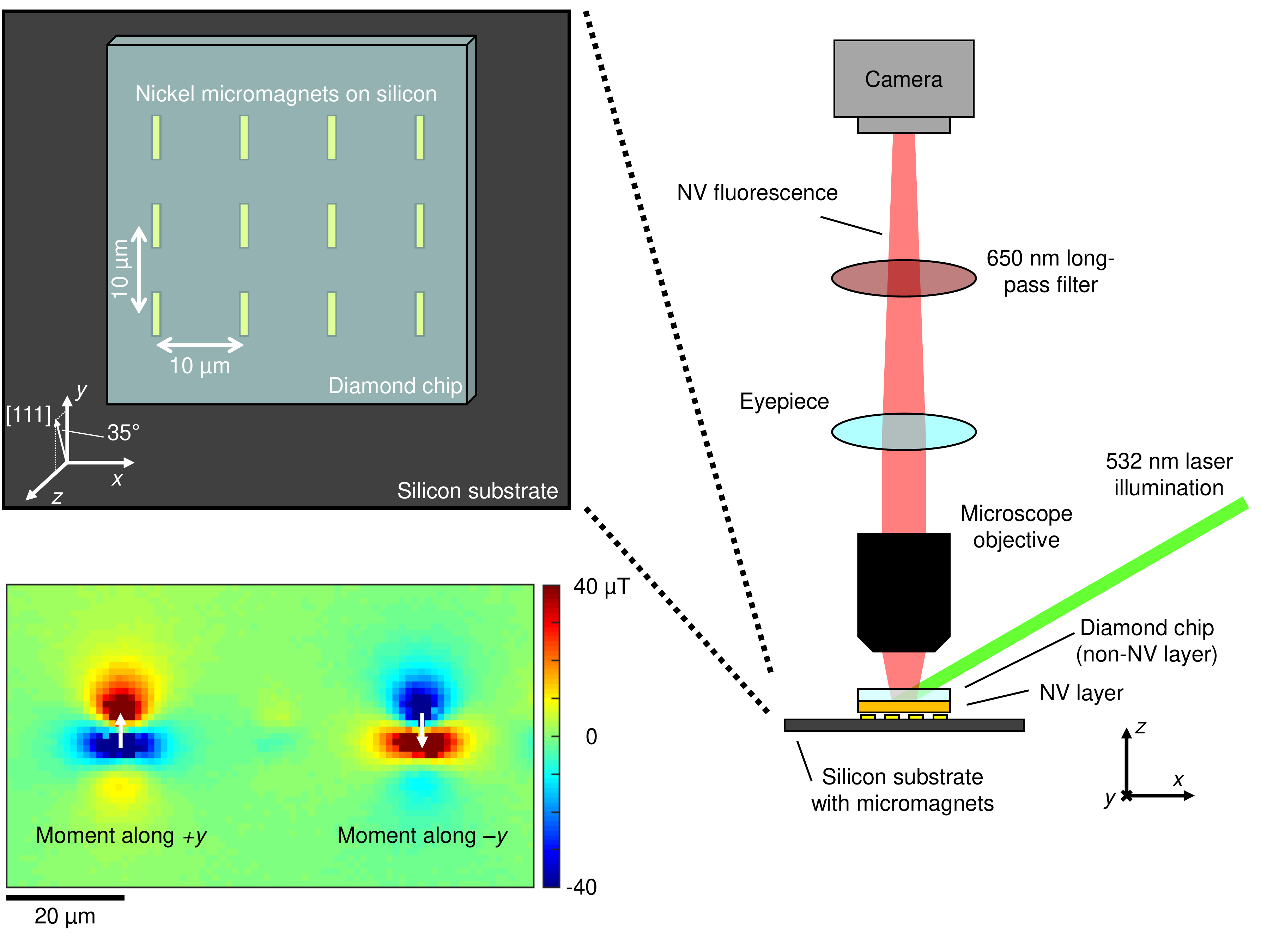}
\put(-2,73){\textsf{\Large a}}
\put(56,73){\textsf{\Large b}}
\put(-2,27){\textsf{\Large c}}
\end{overpic}
\end{center}
\caption{\label{fig1}
(a) A representative layout of a fabricated micromagnet array PUF on silicon, with a diamond chip on top (NV side down) for widefield magnetic imaging. The axes show the coordinate system and the NV [111] crystallographic direction used in this work.
(b) Schematic drawing of a micromagnet array being measured in an NV magnetic imaging setup. A 532 nm pump laser beam illuminates the NV layer, causing them to fluoresce red light that is imaged by a microscope onto a camera sensor. We apply a probe microwave field to interrogate the transition frequencies between NV ground-state magnetic sublevels (microwave wire not shown).
(c) Example $B_{111}$ magnetic field map from two isolated micromagnets, with moments along $+y$ and $-y$ (corresponding to binary 1 and 0 states).
}
\end{figure*}

NV centers are magnetically-sensitive fluorescent defects that are used for quantum magnetic sensing in basic and applied sciences \cite{degenReview, edlynQDMreview, QDM1ggg}. When used to image magnetic fields from sources external to the diamond, NV magnetic imagers offer room-temperature operation, micron-scale spatial resolution (set by the optical diffraction limit), a small sample-sensor distance (as small as a few nanometers), and parallel multi-pixel readout over a few-millimeter field of view. Previous NV widefield magnetic imager experiments have studied the magnetic domains in a hard drive, superconducting and ferromagnetic phase transitions, and current flow in graphene, as well as other applications in geology and biology \cite{hollenbergFilms, rochDAC, heziVortices, tetienneGraphene, QDM1ggg, nv_bacteria, hemozoin}.  Figure \ref{fig1} shows the schematic layout for our NV magnetic imaging experiment. Each micromagnet in the array has an easy magnetic axis that constrains the magnetic moment to be oriented in one of two directions  (along $\pm y$). These micromagnets have random magnetic moment orientations (polarities) at the time of fabrication, which serve as the unique identifier for the PUF. We used nickel as an initial trial material, as nickel micromagnets are simple to fabricate and have a high enough remanence and coercivity to preserve their magnetic moments in few-mT fields. The magnetic field map of an individual micromagnet (measured a few microns away) is well described by a magnetic dipole along the $\pm y$ direction, and we use an NV magnetic imaging setup  to obtain the polarity of each micromagnet in the array.

An NV widefield magnetic imager allows us to measure all micromagnets simultaneously instead of using serial acquisition  with a scanning single-pixel magnetometer, as with a magnetic force microscope (MFM) or a scanning SQUID microscope. Additionally, unlike a microscope based on the magneto-optic Kerr effect (MOKE) which exploits the Faraday rotation of reflected light from a magnetized surface, the NV magnetic imager measures magnetic fields in a plane above the sample. This means the micromagnet PUF can be isolated beneath an opaque layer, protecting the micromagnets from oxidation and making them harder for a counterfeiter to access, while still being readable with an NV magnetic imager. These advantages enable us to read out the micromagnet array quickly and with high sensitivity (as fast as $10^4$ micromagnet states in a few seconds). This work demonstrates that fabricating micromagnet arrays and reading them out with an NV magnetic imager setup is a feasible approach for hardware security, though we note that there is room for additional optimization. Furthermore, this method fulfills the requirements of being straightforward to manufacture, easy to characterize, random, low-cost, and robust.

\begin{figure*}[ht]
\begin{center}
\begin{overpic}[width=0.95\textwidth]{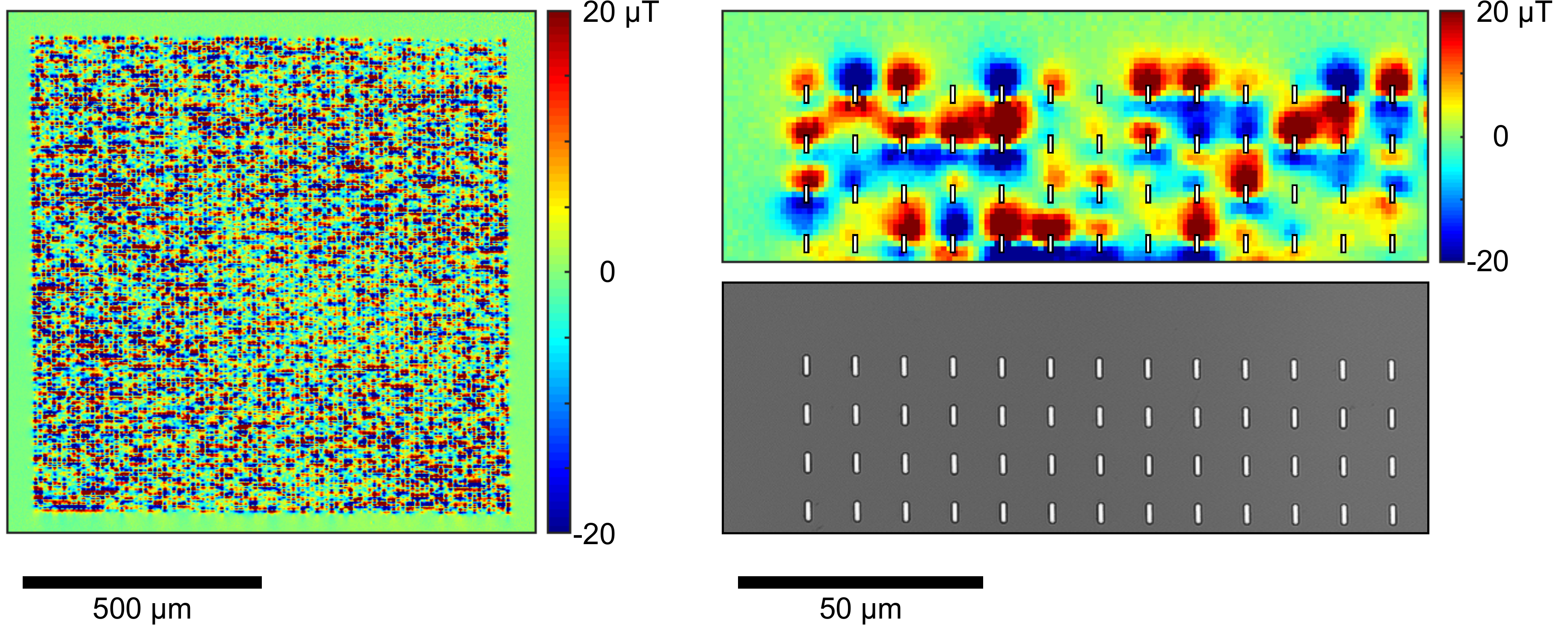}
\put(-2,37){\textsf{\Large a}}
\put(44,37){\textsf{\Large b}}
\put(44,20){\textsf{\Large c}}
\end{overpic}
\end{center}
\caption{\label{examplePUF}
(a) Magnetic image of a micromagnet array (100$\times$100 array with 10 $\upmu$m spacing). To generate this $B_z$ map, we first measured the magnetic field projection onto the NV [111] direction ($B_{111}$), then calculated the $B_z$ map from the $B_{111}$ map \cite{suppl}.
(b) Zoomed-in image of the top-left corner of (a), with the micromagnet locations drawn as an overlay.
(c) Optical image of the same area as in (b).
}
\end{figure*}

\section{Methods}
Our PUF implementation includes three components (Fig.~\ref{fig1}): (i) fabricating the micromagnet array, (ii) measuring the micromagnet polarity states with a widefield magnetic microscopy setup based on NV centers in diamond, and (iii) converting the resulting magnetic image into a binary string (0 or 1 for magnetic polarity along $-y$ or $+y$) using image analysis. The binary string of micromagnet polarities serves as the identifier for the PUF.  In this section we discuss details of the fabrication, readout, and analysis methods we used.

\subsection*{Micromagnet array fabrication}
We fabricated nickel micromagnet arrays on a silicon substrate with 1 $\upmu$m thick thermal oxide.  First, we performed electron-beam lithography using a 30 keV electron beam with 950A7 PMMA as the resist.  After developing the patterns in MIKB/IPA (1:3), we deposited 50 nm of nickel, followed by lift-off. 

Our micromagnet arrays contain $10^4$ micromagnets in a 100$\times$100 grid. The array pitch is 10 $\upmu$m in both the $x$ and $y$ directions (1$\times$1 mm$^2$ total area), and each micromagnet has dimensions of 1$\times$4 $\upmu$m$^2$. Figure \ref{examplePUF} includes magnetic and optical images of a typical micromagnet array. Using bar-shaped micromagnets constrains the magnetic moments along the $\pm y$ direction, which allows us to assign a binary 0 or 1 to each micromagnet moment along $-y$ or $+y$ and simplifies the post-measurement analysis. 

In addition to fabricating uniformly-spaced micromagnet arrays, we also fabricated micromagnets of different sizes and variable spacings (as small as 2 $\upmu$m). We added a 20 nm Al$_2$O$_3$ top layer to protect the nickel micromagnets from oxidation. Analyzing individual micromagnets confirms that the magnetic field map we measure from each micromagnet is well described by a magnetic dipole model and allows us to extract the effective altitude at which the NV layer measures the micromagnets (the standoff distance). Imaging the variable-spaced micromagnets enables us to determine the smallest-resolvable micromagnet spacing. 

\subsection*{Magnetic microscopy readout}
We placed a diamond chip with a 4 $\upmu$m NV layer (grown with isotopically-enriched $^{12}$C abundance) on top of the micromagnet array. By illuminating the NV layer with 532 nm laser light, interrogating the  transition frequencies between the NV ground-state magnetic sublevels with a probe microwave field, and imaging the NV fluorescence intensity with an optical microscope onto a camera sensor, we measured the magnetic field the NVs experience in every pixel. This produces an image of the magnetic field in the NV layer, from which we obtain the polarity of each micromagnet in the array. We minimize the distance between the NV layer and the micromagnet array to maximize the measured field strength and the spatial resolution. 

NV magnetic imaging is an efficient readout tool for micromagnet arrays of this scale because we can image the magnetic field in every pixel simultaneously (rather than raster-scanning a single-pixel detector). Our NV magnetic microscopy instrument has few-micron spatial resolution, 1.09 $\upmu$m pixel size with few-millimeter field-of-view, about 7 $\upmu$T magnetic noise floor in a 1$\times$1 $\upmu$m$^2$ pixel area after 1 second of averaging, and operates in ambient conditions. In this experiment we measured the magnetic field projection $B_{111}$ along the NV [111] crystallographic direction (called ``projection magnetic microscopy" \cite{QDM1ggg}), which points $\sim$35$\degree$ out of the image plane (Fig.~\ref{fig1}a). To simplify the micromagnet array image analysis, we convert the $B_{111}$ map to a $B_z$ map, which is the magnetic field component along the $z$-axis (out of the page) \cite{eduardoUpcont, suppl}.

Figure \ref{examplePUF} includes a representative $B_z$ magnetic map of a micromagnet array measured with our NV magnetic imaging setup in a 20 minute experiment. This measurement duration is chosen to obtain a magnetic map with a signal-to-noise ratio (SNR) of about 100, and could be shortened to a few seconds while still having enough sensitivity to acquire the micromagnet polarities. Since NV magnetic microscopy reads out the magnetic information optically, we can overlay magnetic and white-light optical images to constrain the micromagnet locations.

\subsection*{Magnetic $B_z$ image to bit string conversion}
Given a measured $B_z$ map, we use image analysis techniques from the Python \cite{rossum1995} packages \texttt{scikit-image} \cite{vanderwalt2014} and  \texttt{opencv2} \cite{bradski2000} to analyze the magnetic polarity of each micromagnet in the array. Since the micromagnet array may be slightly rotated in the NV magnetic image, we first align the magnetic array with the coordinate basis of the camera in order to simplify the analysis. This is done by first using Canny edge detection \cite{canny1987} to define edges of regions with and without micromagnet $B_z$ fields, then applying a convex hull-finding algorithm to find the polygon that encloses all of the points found by edge detection.  The slopes of the lines that define the hull indicate the rotation offset of the micromagnet array in the magnetic image. The coordinate bases are aligned by simply rotating the data by this angle (1.3$\degree$ in Fig.~\ref{examplePUF}a).

Once the micromagnet grid is rotated into the $x$-$y$ basis of the image, we identify the $\{x,y\}$ coordinates of each micromagnet by dividing the rectangular area containing the micromagnet fields into a 100$\times$100 grid of cells, each containing the $B_z$ field map from a single micromagnet. Exploiting the fact that the $B_z$ magnetic map for a dipole along the $y$-axis has symmetric positive and negative lobes \cite{suppl}, we sum the $B_z$ field strengths from the top half and the bottom half of each cell, then take the difference, i.e. $\Delta B  = \sum B_{\text{top}} - \sum B_{\text{bottom}}$. If $\Delta B > 0$ the cell is assigned a 1 state for the bit string (moment along $+y$), and if $\Delta B < 0$ the cell is assigned a 0 (moment along $-y$). Using this method, we are able to reliably convert a $B_z$ magnetic map to a bit string for further analysis. This method yields the sign (but not the amplitude) of each dipole moment in the array, though the sign information is sufficient to generate a bit string. 

Our $B_z$ image to bit string conversion method is well suited to this work, due to its simplicity and robustness. We considered several alternative (but more nontrivial) analysis approaches, including fitting every micromagnet to extract the vector moments, using spatial convolutions or matched filters to locate and analyze each micromagnet, and comparing the magnetic field maps from different micromagnet arrays using image comparison software tools (for cases where the micromagnets are too close together to resolve individually).

\begin{figure*}[ht]
\begin{center}
\begin{overpic}[width=0.95\textwidth]{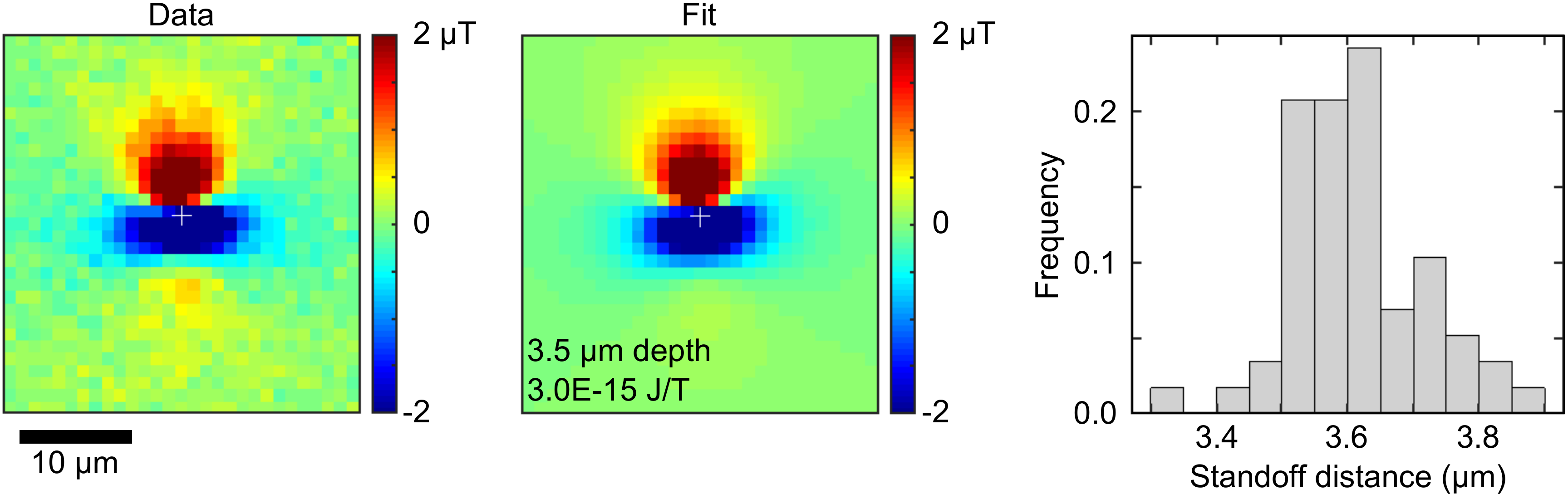}
\put(-1,30){\textsf{\Large a}}
\put(32,30){\textsf{\Large b}}
\put(71,30){\textsf{\Large c}}
\end{overpic}
\end{center}
\caption{\label{standoffAbom}
(a) Example $B_{111}$ data of a 0.25$\times$1 $\upmu$m$^2$ micromagnet measured with a 4 $\upmu$m NV layer.
(b) Magnetic dipole fit for the data in (a).
(c) Histogram of 58 fitted dipole standoff distances, showing a mean standoff of 3.6 $\upmu$m with a 0.1 $\upmu$m standard deviation.
}
\end{figure*}

\section{Results and Discussion}
To understand how to further optimize the micromagnet array PUF and NV readout characteristics, we now assess the NV measurement standoff distance (the effective altitude at which the NV layer measures the micromagnets) and spatial resolution, the NV readout sensitivity, the micromagnet randomness, and the micromagnet coercivity.

\subsection*{Standoff distance and spatial resolution}
For a PUF based on resolving individual micromagnet polarities, the maximum micromagnet areal density is limited by the spatial resolution of the NV magnetic imager setup. The spatial resolution is set by whichever of the following is the largest: (i) the air gap between the NV layer and the micromagnets, (ii) the NV layer thickness, and (iii) the optical diffraction limit \cite{edlynQDMreview}. We aim to minimize the air gap between the NV layer and the magnetic source layer, but this gap can increase if the diamond and sample layers are not flat, or if unwanted dust particles are trapped between them. Similarly, if the NV layer thickness is large compared to the sample layer thickness and the air gap, the microscope will image NV fluorescence from some spatially-averaged NV layer, leading to an increased effective standoff distance. Finally, if the other limitations are minimized, the spatial resolution is set by the optical diffraction limit ($\frac{\lambda}{2 \mathrm{NA}}$, where NA is the numerical aperture and $\lambda$ is the fluorescence wavelength). In our setup we used a microscope objective with NA = 0.25, which corresponds to a 1.4 $\upmu$m diffraction-limited resolution for a typical 700 nm NV fluorescence wavelength, though this could be improved by using an objective lens with a higher NA or by using optical super-resolution techniques \cite{JCsuperres}. We determined that the sample-diamond air gap (about 1.9 $\upmu$m) and the NV layer thickness (4 $\upmu$m) are the main contributors to the standoff distance (3.6 $\upmu$m), and that optical diffraction does not contribute significantly \cite{suppl}.

To arrive at this conclusion, we measured $B_{111}$ magnetic images from isolated micromagnets. Here we used smaller micromagnets (0.25$\times$1 $\upmu$m$^2$) to ensure that we imaged the field from point-like magnetic dipoles (rather than extended sources). We fit the $B_{111}$ magnetic maps from 58 isolated micromagnets using a magnetic dipole model to arrive at a 3.6 $\upmu$m mean standoff distance with this diamond sample (Fig.~\ref{standoffAbom}).

\begin{figure*}[ht]
\begin{center}
\begin{overpic}[width=0.95\textwidth]{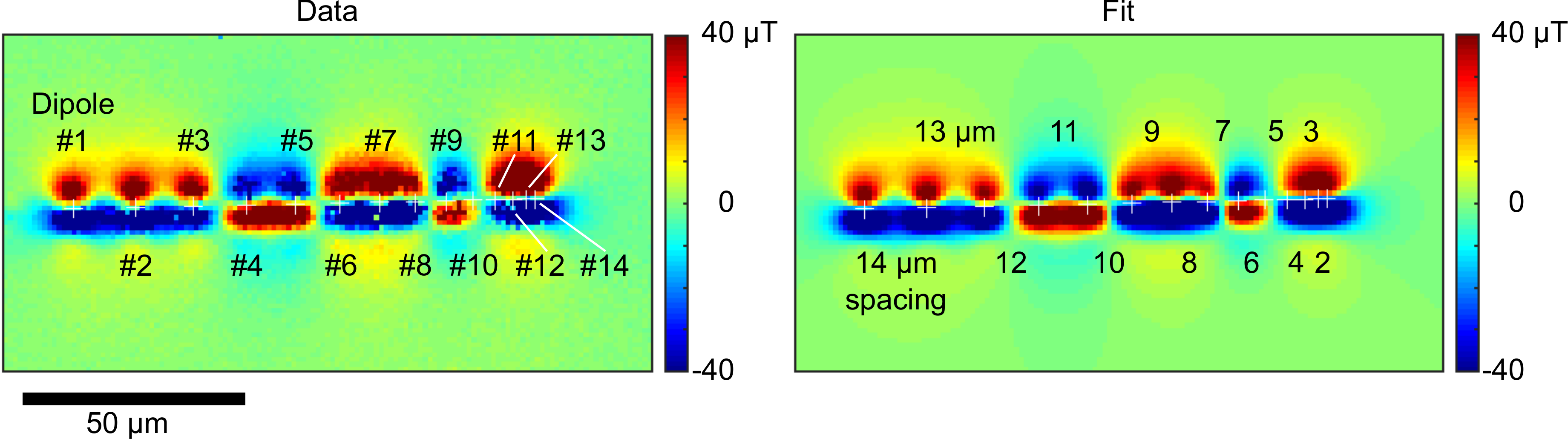}
\put(0,26.5){\textsf{\Large a}}
\put(50.5,26.5){\textsf{\Large b}}
\end{overpic}
\end{center}
\caption{\label{varSpaced} (a) A $B_{111}$ magnetic images of fourteen micromagnets (1$\times$5 $\upmu$m$^2$) with variable spacings between them (14 $\upmu$m to 2 $\upmu$m). 
(b) Fit for the data in (a) using fourteen dipoles with fixed locations.
}
\end{figure*}

From simulations, a 3.6 $\upmu$m standoff distance implies a $\sim$8-10 $\upmu$m minimum micromagnet spacing needed to spatially distinguish neighboring dipoles. However, since the micromagnet moments are constrained along the $y$-axis and the locations are known, we can exploit this information to pack the micromagnets closer together while still being able to determine the state of each one. To see this, Fig.~\ref{varSpaced} shows the magnetic field map from a row of fourteen micromagnets with variable spacings (2 $\upmu$m to 14 $\upmu$m), together with a fourteen-dipole fit with fixed relative locations \cite{nv_bacteria, lennartSQUID}. The magnetic fields from the more closely spaced micromagnets start to blend together, making it difficult to distinguish between whether the source is one micromagnet or several closely-spaced micromagnets. However, given the known micromagnet locations, the fitting routine (and the eye) can distinguish each micromagnet state (for example: dipoles \#9 and \#10). This additional knowledge only helps up to a point; dipoles \#12 -- \#14 are close enough together that the fitting routine is satisfied with modeling these three dipoles as two, assigning all of the magnetic moment to dipoles \#12 and \#14 while setting dipole \#13 to zero. From this example, we conclude that even with a 3.6 $\upmu$m standoff distance, the additional micromagnet  position and moment direction constraints improve this minimum spacing to $\sim$5 $\upmu$m. With the current setup, resolving the moments from micromagnets closer than 2-3 $\upmu$m apart is difficult despite knowing the micromagnet positions.

\subsection*{Magnetic sensitivity and noise floor}
An important criterion for a PUF technology to be useful is that the readout should be reproducible and fast. Reading out a micromagnet array PUF with NV magnetic imaging readout satisfies these requirements. Repeating the measurement shown in Fig.~\ref{examplePUF}a yields the same results. We note that the long-term magnetic moment stability will depend on the properties of the ferromagnetic material, such as the blocking temperature and coercivity.

Despite being able to measure $10^6$ pixels simultaneously with an NV magnetic imager, the SNR in each pixel must be sufficiently high for NV readout to be a practical option. Given that our magnetic sensitivity is about 7 $\upmu$T after 1 second in a 1$\times$1 $\upmu$m$^2$ area, we image the $\sim$20 $\upmu$T typical field strength of the micromagnet array in Fig.~\ref{examplePUF}a with a 0.2 $\upmu$T noise floor (1\%) after about 20 minutes. This SNR and measurement duration could be improved with stronger micromagnet moments, a closer standoff distance, and a better magnetic noise floor.

\subsection{Micromagnet array PUF figure of merit}
For a PUF based on a string of random bits, one figure of merit is the total number of bits. If we interpret this as the bit areal density (bits per mm$^2$), we have demonstrated $10^4$ bits/mm$^2$ in Fig.~\ref{examplePUF}a, though this could be improved to 4\e{4} bits/mm$^2$ with a 5 $\upmu$m micromagnet spacing. This bit areal density is comparable to the bit areal density measured in a 1 GB magnetic hard drive in a previous NV magnetic imaging work (about 5\e{4} bits/mm$^2$) \cite{hollenbergFilms}. Having $10^4$ random bits corresponds to an upper limit of $2^{10000}$ unique identifiers, suggesting that random PUF state duplication is unlikely. This areal density is large enough that one can fabricate a compact micromagnet PUF onto a variety of electronics components as a magnetic tag.

A second interpretation for the figure of merit is the bit readout rate. We measured the magnetic map of 10$^4$ micromagnets in Fig.~\ref{examplePUF}a with SNR = 100 in a 1$\times$1 $\upmu$m$^2$ area after 20 minutes (8 bits/s). This bit rate can be 100$\times$ faster if we relax this to SNR = 10 (a 12 second measurement). Moreover, when converting the $B_z$ map to a bit string, our algorithm takes sums and differences over $\sim$50 $\upmu$m$^2$ areas for each micromagnet. This effectively increases our readout rate by about 7$\times$. Combining these factors yields a 5800 bits/s readout rate. Note that the bit readout rate should be largely independent of the bit areal density, since increasing the areal density also decreases the area and the SNR for each micromagnet.

\subsection*{Micromagnet bit string randomness assessment}

A PUF should be random to provide the best security against potential counterfeiters. This is because even with a perfectly-replicated fabrication process, individual PUF instances are unpredictable and unique, and thus unclonable. To characterize the randomness of a micromagnet PUF bit string, we used a suite of fifteen statistical hypothesis tests designed to test whether a string of binary numbers is consistent with being randomly generated \cite{NISTrng}. Each test reports one or more $p$-value likelihoods of getting such a bit string from a truly random number generator (the null hypothesis), and we reject the null hypothesis for $p < 0.01$.

We applied these tests to the bit string obtained from the measurement in Fig.~\ref{examplePUF}a. The full results are provided in the supplementary information \cite{suppl}. To summarize, we found that this bit string passes most of the statistical tests with a few exceptions: the Approximate Entropy Test, the Runs Test, and some Non-Overlapping Template Test instances. The Approximate Entropy Test examines the entropy of different bit patterns  of length $l$ (calculated as $\sum f_i \ln{f_i}$) for all $2^l$ possible patterns  (each of which appears in the bit string with frequency $f_i$). The Runs Test counts uninterrupted sequences of identical bits (for example, 011110 contains a run of 1's with length 4) and indicates whether the values in the bit string are fluctuating too quickly or too slowly. Our bit string failed because there were too few runs, suggesting that the micromagnet polarities fluctuate too slowly. The Non-Overlapping Template Test searches for how often different 9-bit patterns occur to see if these patterns appear too frequently.  Our bit string passed with 138 patterns out of 148. Note that some of the statistical tests (such as the Random Excursions Test) require a longer bit string than we provided, and were therefore not evaluated.

Despite the partial success of the randomness assessment above, this outcome is better than expected for this first attempt. The micromagnets were prepared in the Earth's magnetic field ($B_\textrm{e} \approx$ 50 $\upmu$T), meaning that naively they should be preferentially aligned with the Earth's field (or other magnetic fields in the lab) during fabrication.  Even the weakly-magnetized 0.25$\times$1 $\upmu$m$^2$ micromagnet shown in Fig.~\ref{standoffAbom}a ($m = 3.0\e{-15}$ J/T moment) has a large energy in the Earth's field ($m B_\textrm{e} = 1.5\e{-19}$ J) compared to the Boltzmann energy ($k_\textrm{B} T = 4.1\e{-21}$ J at 300 K).  Furthermore, each micromagnet experiences the magnetic fields from its neighbors, leading to the micromagnet moment states potentially being correlated between neighbors in an antiferromagnetic pattern. Although these potential pitfalls may affect the randomness of the bit string we extract from a micromagnet array PUF, fortunately the example micromagnet array in Fig.~\ref{examplePUF}a still passes most of the randomness tests without additional processing steps.

\subsection*{Micromagnet coercivity and remanence}
NV magnetic sensing is usually done at nonzero bias field (1 mT or higher), which can pose a problem if this field is strong enough to remagnetize the PUF micromagnets.  Furthermore, in order to preserve the magnetic moments from the outside magnetic or thermal environment, the micromagnets should be single-domain ferromagnetic (rather than superparamagnetic or multidomain). The micromagnet material, aspect ratio, and size determine its magnetic coercivity and remanence, both of which we aim to maximize.

To verify that the choice of micromagnet specifications suit our requirements, we measured the individual micromagnets with equal and opposite bias fields ($\pm$1.3 mT along the [111] direction) and compared the resulting magnetic moments \cite{QDM1ggg}. Although the moments varied by a few tens of percent between the measurements, the polarities (which are used as the PUF identifier) were unaffected. Note that fabricating micromagnets with different materials (SmCo, for instance) may yield better coercivity and remanence than we get with nickel. 

\subsection*{Overcoming current limitations}
The idea of a micromagnet-based PUF has some drawbacks independent of the particular implementation details. A counterfeiter might try to copy a micromagnet array PUF by individually magnetizing each micromagnet with a hard-drive read-write head or an MFM tip. One can make this impractical by packing the micromagnets closer together, fabricating more of them in the array, or isolating them beneath a nonmagnetic protective layer (making it difficult to address a particular micromagnet without affecting its neighbors). 

Compared to other PUF implementations, our micromagnet array PUF is categorized as a ``weak PUF", since it provides one response state (the magnetic moment bit string) when measured \cite{PUFtaxonomy}. A ``strong PUF" has more security by having a large database of possible measurements and responses. One could adapt the current micromagnet PUF into being a strong PUF by making magnetic hysteresis or first order reversal curve (FORC) measurements, sampling from a large list of bias magnetic fields. 

\subsection*{Further optimization}
One direction for further improvement is to optimize the micromagnet density and the readout SNR. We can vary the micromagnet dimensions and spacing, the micromagnet material, and the NV layer thickness. By varying these quantities, we must balance the resulting tradeoffs in the standoff distance, magnetic moment, measured field strength, and single-pixel magnetic noise floor. For example, increasing the NV layer thickness increases the NV fluorescence intensity (resulting in better magnetic sensitivity) but can also increase the standoff distance (resulting in a weaker magnetic field strength). Packing the micromagnets closer together increases the areal density, but could also require us to use a thinner NV layer with better spatial resolution but worse magnetic noise floor.

To quantify their resilience to the environment, one can expose the micromagnet array PUF to a range of temperatures and magnetic fields to evaluate the magnetic coercivity, blocking temperature, and possible chemical reactions as the micromagnets age or heat up. Knowing these specifications is necessary if micromagnet array PUFs become a more standard anti-counterfeiting technique, especially if they are used in extreme environments. Similarly, one might verify that alternating-field or thermal demagnetization can erase and reinitialize a micromagnet PUF to a new random state \cite{tauxePaleomag}.

There are materials better suited for micromagnet array PUFs than nickel. First, nickel is susceptible to oxidation, and other materials may resist viscous remanent magnetization more effectively \cite{elements209}. Furthermore, the micromagnet moments in the current fabrication method are not uniform, as seen in Fig.~\ref{examplePUF}a. This could present a problem if the fields from weaker micromagnets are overwhelmed by the fields from stronger ones nearby, which can complicate the post-measurement analysis (especially when packing the micromagnets close together). To further optimize the moment uniformity, coercivity, and remanence, one may experiment with different micromagnet dimensions and materials, and image the micromagnet hysteresis curves.

\section{Conclusion and Outlook}

In this  work we demonstrated  the idea of using NV magnetic microscopy as a readout technique for an array of fabricated micromagnets. These components, together with post-measurement image analysis, constitute a PUF for hardware security and trust validation. A micromagnet array PUF is straightforward to fabricate and measure, has random magnetic polarities, and is robust against the few-mT fields used in the NV magnetic imaging setup. To further validate that each micromagnet PUF is unique, we must measure and compare many instances to confirm that the inter-device separation is sufficiently large. 

A micromagnet array PUF has some similarities to a magnetic stripe PUF \cite{magneticStripePatent1, magneticStripePatent2}, though our implementation benefits from being more compact, having more bits, being CMOS-compatible, being able to read magnetic moments in arbitrary directions, and being more difficult to copy.  In particular, one can use a micromagnet array PUF to tag sensitive application-specific integrated circuits (ASICs) for counterfeit protection. Conversely, unlike an electronic PUF, a micromagnet array PUF could tag unpowered devices. One could also use a micromagnet array PUF as a physical cryptographic key for situations where storing the key electronically is undesirable \cite{rockOfRandomness}. A PUF readout should be repeatable on short and long timescales; a micromagnet array PUF may offer favorable long-term stability compared to other approaches, as magnetic domains in geological samples can last for billions of years.

A magnetic PUF implementation could also work with other magnetic readout schemes besides NV magnetic imaging (such as scanning SQUID microscopy, MFM, and MOKE microscopy) provided that the micromagnet spacing and moments match the spatial resolution and moment sensitivity. However, using some of these tools would sacrifice  the widefield parallel readout and the ability to sense the micromagnets below a nonmagnetic protective layer that the NV magnetic imager provides. We fabricated bar-shaped micromagnets for straightforward analysis and characterization. One could instead fabricate disk-shaped micromagnets for greater PUF randomness, since the magnetic moment vectors can point in any direction in the $x$-$y$  plane rather than being constrained along the $y$-axis. Similarly, if we could fabricate micromagnets with moments along the $z$-axis, one could pack them closer together and simplify the magnetic-to-bit string analysis compared to a $y$-aligned micromagnet array.

\section{Acknowledgements}
We thank Todd Bauer, Jason Hamlet, Jacob Henshaw, Yuan-Yu Jau, and Peter Schwindt for providing feedback on the manuscript, and Michael Lilly for help in the clean room. Sandia National Laboratories is a multi-mission laboratory managed and operated by National Technology and Engineering Solutions of Sandia, LLC, a wholly owned subsidiary of Honeywell International, Inc., for the DOE's National Nuclear Security Administration under contract DE-NA0003525. This work was funded by the Laboratory Directed Research and Development Program and performed, in part, at the Center for Integrated Nanotechnologies, an Office of Science User Facility operated for the U.S.~Department of Energy (DOE) Office of Science. P.K.~is supported by the Sandia National Laboratories Truman Fellowship Program.  

\begin{thebibliography}{29}%
\makeatletter
\providecommand \@ifxundefined [1]{%
 \@ifx{#1\undefined}
}%
\providecommand \@ifnum [1]{%
 \ifnum #1\expandafter \@firstoftwo
 \else \expandafter \@secondoftwo
 \fi
}%
\providecommand \@ifx [1]{%
 \ifx #1\expandafter \@firstoftwo
 \else \expandafter \@secondoftwo
 \fi
}%
\providecommand \natexlab [1]{#1}%
\providecommand \enquote  [1]{``#1''}%
\providecommand \bibnamefont  [1]{#1}%
\providecommand \bibfnamefont [1]{#1}%
\providecommand \citenamefont [1]{#1}%
\providecommand \href@noop [0]{\@secondoftwo}%
\providecommand \href [0]{\begingroup \@sanitize@url \@href}%
\providecommand \@href[1]{\@@startlink{#1}\@@href}%
\providecommand \@@href[1]{\endgroup#1\@@endlink}%
\providecommand \@sanitize@url [0]{\catcode `\\12\catcode `\$12\catcode
  `\&12\catcode `\#12\catcode `\^12\catcode `\_12\catcode `\%12\relax}%
\providecommand \@@startlink[1]{}%
\providecommand \@@endlink[0]{}%
\providecommand \url  [0]{\begingroup\@sanitize@url \@url }%
\providecommand \@url [1]{\endgroup\@href {#1}{\urlprefix }}%
\providecommand \urlprefix  [0]{URL }%
\providecommand \Eprint [0]{\href }%
\providecommand \doibase [0]{http://dx.doi.org/}%
\providecommand \selectlanguage [0]{\@gobble}%
\providecommand \bibinfo  [0]{\@secondoftwo}%
\providecommand \bibfield  [0]{\@secondoftwo}%
\providecommand \translation [1]{[#1]}%
\providecommand \BibitemOpen [0]{}%
\providecommand \bibitemStop [0]{}%
\providecommand \bibitemNoStop [0]{.\EOS\space}%
\providecommand \EOS [0]{\spacefactor3000\relax}%
\providecommand \BibitemShut  [1]{\csname bibitem#1\endcsname}%
\let\auto@bib@innerbib\@empty
\bibitem [{\citenamefont {Maes}(2013)}]{maesPUFBook}%
  \BibitemOpen
  \bibfield  {author} {\bibinfo {author} {\bibfnamefont {R.}~\bibnamefont
  {Maes}},\ }\href@noop {} {\emph {\bibinfo {title} {Physically {U}nclonable
  {F}unctions: {C}onstructions, {P}roperties and {A}pplications}}}\ (\bibinfo
  {publisher} {Springer-Verlag},\ \bibinfo {year} {2013})\BibitemShut {NoStop}%
\bibitem [{\citenamefont {Herder}\ \emph {et~al.}(2014)\citenamefont {Herder},
  \citenamefont {Yu}, \citenamefont {Koushanfar},\ and\ \citenamefont
  {Devadas}}]{herder2014}%
  \BibitemOpen
  \bibfield  {author} {\bibinfo {author} {\bibfnamefont {C.}~\bibnamefont
  {Herder}}, \bibinfo {author} {\bibfnamefont {M.-D.}\ \bibnamefont {Yu}},
  \bibinfo {author} {\bibfnamefont {F.}~\bibnamefont {Koushanfar}}, \ and\
  \bibinfo {author} {\bibfnamefont {S.}~\bibnamefont {Devadas}},\ }\href@noop
  {} {\bibfield  {journal} {\bibinfo  {journal} {Proceedings of the IEEE}\
  }\textbf {\bibinfo {volume} {102}},\ \bibinfo {pages} {1126} (\bibinfo {year}
  {2014})}\BibitemShut {NoStop}%
\bibitem [{\citenamefont {Katzenbeisser}\ \emph {et~al.}(2012)\citenamefont
  {Katzenbeisser}, \citenamefont {Kocaba{\c{s}}}, \citenamefont
  {Ro{\v{z}}i{\'c}}, \citenamefont {Sadeghi}, \citenamefont {Verbauwhede},\
  and\ \citenamefont {Wachsmann}}]{katzenbeisser2012}%
  \BibitemOpen
  \bibfield  {author} {\bibinfo {author} {\bibfnamefont {S.}~\bibnamefont
  {Katzenbeisser}}, \bibinfo {author} {\bibfnamefont {{\"U}.}~\bibnamefont
  {Kocaba{\c{s}}}}, \bibinfo {author} {\bibfnamefont {V.}~\bibnamefont
  {Ro{\v{z}}i{\'c}}}, \bibinfo {author} {\bibfnamefont {A.-R.}\ \bibnamefont
  {Sadeghi}}, \bibinfo {author} {\bibfnamefont {I.}~\bibnamefont
  {Verbauwhede}}, \ and\ \bibinfo {author} {\bibfnamefont {C.}~\bibnamefont
  {Wachsmann}},\ }in\ \href@noop {} {\emph {\bibinfo {booktitle} {International
  Workshop on Cryptographic Hardware and Embedded Systems}}}\ (\bibinfo
  {organization} {Springer},\ \bibinfo {year} {2012})\ pp.\ \bibinfo {pages}
  {283--301}\BibitemShut {NoStop}%
\bibitem [{\citenamefont {Lim}\ \emph {et~al.}(2005)\citenamefont {Lim},
  \citenamefont {Lee}, \citenamefont {Gassend}, \citenamefont {Suh},
  \citenamefont {Van~Dijk},\ and\ \citenamefont {Devadas}}]{lim2005}%
  \BibitemOpen
  \bibfield  {author} {\bibinfo {author} {\bibfnamefont {D.}~\bibnamefont
  {Lim}}, \bibinfo {author} {\bibfnamefont {J.~W.}\ \bibnamefont {Lee}},
  \bibinfo {author} {\bibfnamefont {B.}~\bibnamefont {Gassend}}, \bibinfo
  {author} {\bibfnamefont {G.~E.}\ \bibnamefont {Suh}}, \bibinfo {author}
  {\bibfnamefont {M.}~\bibnamefont {Van~Dijk}}, \ and\ \bibinfo {author}
  {\bibfnamefont {S.}~\bibnamefont {Devadas}},\ }\href@noop {} {\bibfield
  {journal} {\bibinfo  {journal} {IEEE Transactions on Very Large Scale
  Integration (VLSI) Systems}\ }\textbf {\bibinfo {volume} {13}},\ \bibinfo
  {pages} {1200} (\bibinfo {year} {2005})}\BibitemShut {NoStop}%
\bibitem [{\citenamefont {McGrath}\ \emph {et~al.}(2019)\citenamefont
  {McGrath}, \citenamefont {Bagci}, \citenamefont {Wang}, \citenamefont
  {Roedig},\ and\ \citenamefont {Young}}]{PUFtaxonomy}%
  \BibitemOpen
  \bibfield  {author} {\bibinfo {author} {\bibfnamefont {T.}~\bibnamefont
  {McGrath}}, \bibinfo {author} {\bibfnamefont {I.~E.}\ \bibnamefont {Bagci}},
  \bibinfo {author} {\bibfnamefont {Z.~M.}\ \bibnamefont {Wang}}, \bibinfo
  {author} {\bibfnamefont {U.}~\bibnamefont {Roedig}}, \ and\ \bibinfo {author}
  {\bibfnamefont {R.~J.}\ \bibnamefont {Young}},\ }\href {\doibase
  10.1063/1.5079407} {\bibfield  {journal} {\bibinfo  {journal} {Applied
  Physics Reviews}\ }\textbf {\bibinfo {volume} {6}},\ \bibinfo {pages}
  {011303} (\bibinfo {year} {2019})}\BibitemShut {NoStop}%
\bibitem [{\citenamefont {Bauer}\ and\ \citenamefont
  {Hamlet}(2014)}]{sandiaPUFsPrimer}%
  \BibitemOpen
  \bibfield  {author} {\bibinfo {author} {\bibfnamefont {T.}~\bibnamefont
  {Bauer}}\ and\ \bibinfo {author} {\bibfnamefont {J.}~\bibnamefont {Hamlet}},\
  }\href {\doibase 10.1109/MSP.2014.123} {\bibfield  {journal} {\bibinfo
  {journal} {IEEE Security \& Privacy}\ }\textbf {\bibinfo {volume} {12}},\
  \bibinfo {pages} {97} (\bibinfo {year} {2014})}\BibitemShut {NoStop}%
\bibitem [{\citenamefont {Schirhagl}\ \emph {et~al.}(2014)\citenamefont
  {Schirhagl}, \citenamefont {Chang}, \citenamefont {Loretz},\ and\
  \citenamefont {Degen}}]{degenReview}%
  \BibitemOpen
  \bibfield  {author} {\bibinfo {author} {\bibfnamefont {R.}~\bibnamefont
  {Schirhagl}}, \bibinfo {author} {\bibfnamefont {K.}~\bibnamefont {Chang}},
  \bibinfo {author} {\bibfnamefont {M.}~\bibnamefont {Loretz}}, \ and\ \bibinfo
  {author} {\bibfnamefont {C.~L.}\ \bibnamefont {Degen}},\ }\href {\doibase
  10.1146/annurev-physchem-040513-103659} {\bibfield  {journal} {\bibinfo
  {journal} {Annual Review of Physical Chemistry}\ }\textbf {\bibinfo {volume}
  {65}},\ \bibinfo {pages} {83} (\bibinfo {year} {2014})}\BibitemShut {NoStop}%
\bibitem [{\citenamefont {Levine}\ \emph {et~al.}(2019)\citenamefont {Levine},
  \citenamefont {Turner}, \citenamefont {Kehayias}, \citenamefont {Hart},
  \citenamefont {Langellier}, \citenamefont {Trubko}, \citenamefont {Glenn},
  \citenamefont {Fu},\ and\ \citenamefont {Walsworth}}]{edlynQDMreview}%
  \BibitemOpen
  \bibfield  {author} {\bibinfo {author} {\bibfnamefont {E.~V.}\ \bibnamefont
  {Levine}}, \bibinfo {author} {\bibfnamefont {M.~J.}\ \bibnamefont {Turner}},
  \bibinfo {author} {\bibfnamefont {P.}~\bibnamefont {Kehayias}}, \bibinfo
  {author} {\bibfnamefont {C.~A.}\ \bibnamefont {Hart}}, \bibinfo {author}
  {\bibfnamefont {N.}~\bibnamefont {Langellier}}, \bibinfo {author}
  {\bibfnamefont {R.}~\bibnamefont {Trubko}}, \bibinfo {author} {\bibfnamefont
  {D.~R.}\ \bibnamefont {Glenn}}, \bibinfo {author} {\bibfnamefont {R.~R.}\
  \bibnamefont {Fu}}, \ and\ \bibinfo {author} {\bibfnamefont {R.~L.}\
  \bibnamefont {Walsworth}},\ }\href@noop {} {\bibfield  {journal} {\bibinfo
  {journal} {Nanophotonics}\ }\textbf {\bibinfo {volume} {8}},\ \bibinfo
  {pages} {1945} (\bibinfo {year} {2019})}\BibitemShut {NoStop}%
\bibitem [{\citenamefont {Glenn}\ \emph {et~al.}(2017)\citenamefont {Glenn},
  \citenamefont {Fu}, \citenamefont {Kehayias}, \citenamefont {Le~Sage},
  \citenamefont {Lima}, \citenamefont {Weiss},\ and\ \citenamefont
  {Walsworth}}]{QDM1ggg}%
  \BibitemOpen
  \bibfield  {author} {\bibinfo {author} {\bibfnamefont {D.~R.}\ \bibnamefont
  {Glenn}}, \bibinfo {author} {\bibfnamefont {R.~R.}\ \bibnamefont {Fu}},
  \bibinfo {author} {\bibfnamefont {P.}~\bibnamefont {Kehayias}}, \bibinfo
  {author} {\bibfnamefont {D.}~\bibnamefont {Le~Sage}}, \bibinfo {author}
  {\bibfnamefont {E.~A.}\ \bibnamefont {Lima}}, \bibinfo {author}
  {\bibfnamefont {B.~P.}\ \bibnamefont {Weiss}}, \ and\ \bibinfo {author}
  {\bibfnamefont {R.~L.}\ \bibnamefont {Walsworth}},\ }\href {\doibase
  10.1002/2017GC006946} {\bibfield  {journal} {\bibinfo  {journal}
  {Geochemistry, Geophysics, Geosystems}\ }\textbf {\bibinfo {volume} {18}},\
  \bibinfo {pages} {3254} (\bibinfo {year} {2017})}\BibitemShut {NoStop}%
\bibitem [{\citenamefont {Simpson}\ \emph {et~al.}(2016)\citenamefont
  {Simpson}, \citenamefont {Tetienne}, \citenamefont {McCoey}, \citenamefont
  {Ganesan}, \citenamefont {Hall}, \citenamefont {Petrou}, \citenamefont
  {Scholten},\ and\ \citenamefont {Hollenberg}}]{hollenbergFilms}%
  \BibitemOpen
  \bibfield  {author} {\bibinfo {author} {\bibfnamefont {D.~A.}\ \bibnamefont
  {Simpson}}, \bibinfo {author} {\bibfnamefont {J.-P.}\ \bibnamefont
  {Tetienne}}, \bibinfo {author} {\bibfnamefont {J.~M.}\ \bibnamefont
  {McCoey}}, \bibinfo {author} {\bibfnamefont {K.}~\bibnamefont {Ganesan}},
  \bibinfo {author} {\bibfnamefont {L.~T.}\ \bibnamefont {Hall}}, \bibinfo
  {author} {\bibfnamefont {S.}~\bibnamefont {Petrou}}, \bibinfo {author}
  {\bibfnamefont {R.~E.}\ \bibnamefont {Scholten}}, \ and\ \bibinfo {author}
  {\bibfnamefont {L.~C.}\ \bibnamefont {Hollenberg}},\ }\href@noop {}
  {\bibfield  {journal} {\bibinfo  {journal} {Scientific reports}\ }\textbf
  {\bibinfo {volume} {6}},\ \bibinfo {pages} {1} (\bibinfo {year}
  {2016})}\BibitemShut {NoStop}%
\bibitem [{\citenamefont {Lesik}\ \emph {et~al.}(2019)\citenamefont {Lesik},
  \citenamefont {Plisson}, \citenamefont {Toraille}, \citenamefont {Renaud},
  \citenamefont {Occelli}, \citenamefont {Schmidt}, \citenamefont {Salord},
  \citenamefont {Delobbe}, \citenamefont {Debuisschert}, \citenamefont
  {Rondin}, \citenamefont {Loubeyre},\ and\ \citenamefont {Roch}}]{rochDAC}%
  \BibitemOpen
  \bibfield  {author} {\bibinfo {author} {\bibfnamefont {M.}~\bibnamefont
  {Lesik}}, \bibinfo {author} {\bibfnamefont {T.}~\bibnamefont {Plisson}},
  \bibinfo {author} {\bibfnamefont {L.}~\bibnamefont {Toraille}}, \bibinfo
  {author} {\bibfnamefont {J.}~\bibnamefont {Renaud}}, \bibinfo {author}
  {\bibfnamefont {F.}~\bibnamefont {Occelli}}, \bibinfo {author} {\bibfnamefont
  {M.}~\bibnamefont {Schmidt}}, \bibinfo {author} {\bibfnamefont
  {O.}~\bibnamefont {Salord}}, \bibinfo {author} {\bibfnamefont
  {A.}~\bibnamefont {Delobbe}}, \bibinfo {author} {\bibfnamefont
  {T.}~\bibnamefont {Debuisschert}}, \bibinfo {author} {\bibfnamefont
  {L.}~\bibnamefont {Rondin}}, \bibinfo {author} {\bibfnamefont
  {P.}~\bibnamefont {Loubeyre}}, \ and\ \bibinfo {author} {\bibfnamefont
  {J.-F.}\ \bibnamefont {Roch}},\ }\href {\doibase 10.1126/science.aaw4329}
  {\bibfield  {journal} {\bibinfo  {journal} {Science}\ }\textbf {\bibinfo
  {volume} {366}},\ \bibinfo {pages} {1359} (\bibinfo {year}
  {2019})}\BibitemShut {NoStop}%
\bibitem [{\citenamefont {Schlussel}\ \emph {et~al.}(2018)\citenamefont
  {Schlussel}, \citenamefont {Lenz}, \citenamefont {Rohner}, \citenamefont
  {Bar-Haim}, \citenamefont {Bougas}, \citenamefont {Groswasser}, \citenamefont
  {Kieschnick}, \citenamefont {Rozenberg}, \citenamefont {Thiel}, \citenamefont
  {Waxman}, \citenamefont {Meijer}, \citenamefont {Maletinsky}, \citenamefont
  {Budker},\ and\ \citenamefont {Folman}}]{heziVortices}%
  \BibitemOpen
  \bibfield  {author} {\bibinfo {author} {\bibfnamefont {Y.}~\bibnamefont
  {Schlussel}}, \bibinfo {author} {\bibfnamefont {T.}~\bibnamefont {Lenz}},
  \bibinfo {author} {\bibfnamefont {D.}~\bibnamefont {Rohner}}, \bibinfo
  {author} {\bibfnamefont {Y.}~\bibnamefont {Bar-Haim}}, \bibinfo {author}
  {\bibfnamefont {L.}~\bibnamefont {Bougas}}, \bibinfo {author} {\bibfnamefont
  {D.}~\bibnamefont {Groswasser}}, \bibinfo {author} {\bibfnamefont
  {M.}~\bibnamefont {Kieschnick}}, \bibinfo {author} {\bibfnamefont
  {E.}~\bibnamefont {Rozenberg}}, \bibinfo {author} {\bibfnamefont
  {L.}~\bibnamefont {Thiel}}, \bibinfo {author} {\bibfnamefont
  {A.}~\bibnamefont {Waxman}}, \bibinfo {author} {\bibfnamefont
  {J.}~\bibnamefont {Meijer}}, \bibinfo {author} {\bibfnamefont
  {P.}~\bibnamefont {Maletinsky}}, \bibinfo {author} {\bibfnamefont
  {D.}~\bibnamefont {Budker}}, \ and\ \bibinfo {author} {\bibfnamefont
  {R.}~\bibnamefont {Folman}},\ }\href {\doibase
  10.1103/PhysRevApplied.10.034032} {\bibfield  {journal} {\bibinfo  {journal}
  {Phys. Rev. Applied}\ }\textbf {\bibinfo {volume} {10}},\ \bibinfo {pages}
  {034032} (\bibinfo {year} {2018})}\BibitemShut {NoStop}%
\bibitem [{\citenamefont {Tetienne}\ \emph {et~al.}(2017)\citenamefont
  {Tetienne}, \citenamefont {Dontschuk}, \citenamefont {Broadway},
  \citenamefont {Stacey}, \citenamefont {Simpson},\ and\ \citenamefont
  {Hollenberg}}]{tetienneGraphene}%
  \BibitemOpen
  \bibfield  {author} {\bibinfo {author} {\bibfnamefont {J.-P.}\ \bibnamefont
  {Tetienne}}, \bibinfo {author} {\bibfnamefont {N.}~\bibnamefont {Dontschuk}},
  \bibinfo {author} {\bibfnamefont {D.~A.}\ \bibnamefont {Broadway}}, \bibinfo
  {author} {\bibfnamefont {A.}~\bibnamefont {Stacey}}, \bibinfo {author}
  {\bibfnamefont {D.~A.}\ \bibnamefont {Simpson}}, \ and\ \bibinfo {author}
  {\bibfnamefont {L.~C.~L.}\ \bibnamefont {Hollenberg}},\ }\href {\doibase
  10.1126/sciadv.1602429} {\bibfield  {journal} {\bibinfo  {journal} {Science
  Advances}\ }\textbf {\bibinfo {volume} {3}} (\bibinfo {year} {2017}),\
  10.1126/sciadv.1602429}\BibitemShut {NoStop}%
\bibitem [{\citenamefont {Le~Sage}\ \emph {et~al.}(2013)\citenamefont
  {Le~Sage}, \citenamefont {Arai}, \citenamefont {Glenn}, \citenamefont
  {DeVience}, \citenamefont {Pham}, \citenamefont {Rahn-Lee}, \citenamefont
  {Lukin}, \citenamefont {Yacoby}, \citenamefont {Komeili},\ and\ \citenamefont
  {Walsworth}}]{nv_bacteria}%
  \BibitemOpen
  \bibfield  {author} {\bibinfo {author} {\bibfnamefont {D.}~\bibnamefont
  {Le~Sage}}, \bibinfo {author} {\bibfnamefont {K.}~\bibnamefont {Arai}},
  \bibinfo {author} {\bibfnamefont {D.~R.}\ \bibnamefont {Glenn}}, \bibinfo
  {author} {\bibfnamefont {S.~J.}\ \bibnamefont {DeVience}}, \bibinfo {author}
  {\bibfnamefont {L.~M.}\ \bibnamefont {Pham}}, \bibinfo {author}
  {\bibfnamefont {L.}~\bibnamefont {Rahn-Lee}}, \bibinfo {author}
  {\bibfnamefont {M.~D.}\ \bibnamefont {Lukin}}, \bibinfo {author}
  {\bibfnamefont {A.}~\bibnamefont {Yacoby}}, \bibinfo {author} {\bibfnamefont
  {A.}~\bibnamefont {Komeili}}, \ and\ \bibinfo {author} {\bibfnamefont
  {R.~L.}\ \bibnamefont {Walsworth}},\ }\href
  {http://dx.doi.org/10.1038/nature12072} {\bibfield  {journal} {\bibinfo
  {journal} {Nature}\ }\textbf {\bibinfo {volume} {496}},\ \bibinfo {pages}
  {486} (\bibinfo {year} {2013})}\BibitemShut {NoStop}%
\bibitem [{\citenamefont {Fescenko}\ \emph {et~al.}(2019)\citenamefont
  {Fescenko}, \citenamefont {Laraoui}, \citenamefont {Smits}, \citenamefont
  {Mosavian}, \citenamefont {Kehayias}, \citenamefont {Seto}, \citenamefont
  {Bougas}, \citenamefont {Jarmola},\ and\ \citenamefont {Acosta}}]{hemozoin}%
  \BibitemOpen
  \bibfield  {author} {\bibinfo {author} {\bibfnamefont {I.}~\bibnamefont
  {Fescenko}}, \bibinfo {author} {\bibfnamefont {A.}~\bibnamefont {Laraoui}},
  \bibinfo {author} {\bibfnamefont {J.}~\bibnamefont {Smits}}, \bibinfo
  {author} {\bibfnamefont {N.}~\bibnamefont {Mosavian}}, \bibinfo {author}
  {\bibfnamefont {P.}~\bibnamefont {Kehayias}}, \bibinfo {author}
  {\bibfnamefont {J.}~\bibnamefont {Seto}}, \bibinfo {author} {\bibfnamefont
  {L.}~\bibnamefont {Bougas}}, \bibinfo {author} {\bibfnamefont
  {A.}~\bibnamefont {Jarmola}}, \ and\ \bibinfo {author} {\bibfnamefont
  {V.~M.}\ \bibnamefont {Acosta}},\ }\href {\doibase
  10.1103/PhysRevApplied.11.034029} {\bibfield  {journal} {\bibinfo  {journal}
  {Phys. Rev. Applied}\ }\textbf {\bibinfo {volume} {11}},\ \bibinfo {pages}
  {034029} (\bibinfo {year} {2019})}\BibitemShut {NoStop}%
\bibitem [{sup()}]{suppl}%
  \BibitemOpen
  \href@noop {} {}\bibinfo {note} {Additional details are included in the
  supplemental material.}\BibitemShut {Stop}%
\bibitem [{\citenamefont {Lima}\ and\ \citenamefont
  {Weiss}(2009)}]{eduardoUpcont}%
  \BibitemOpen
  \bibfield  {author} {\bibinfo {author} {\bibfnamefont {E.~A.}\ \bibnamefont
  {Lima}}\ and\ \bibinfo {author} {\bibfnamefont {B.~P.}\ \bibnamefont
  {Weiss}},\ }\href@noop {} {\bibfield  {journal} {\bibinfo  {journal} {Journal
  of Geophysical Research: Solid Earth}\ }\textbf {\bibinfo {volume} {114}}
  (\bibinfo {year} {2009})}\BibitemShut {NoStop}%
\bibitem [{\citenamefont {Rossum}(1995)}]{rossum1995}%
  \BibitemOpen
  \bibfield  {author} {\bibinfo {author} {\bibfnamefont {G.}~\bibnamefont
  {Rossum}},\ }\href@noop {} {\emph {\bibinfo {title} {Python Reference
  Manual}}},\ \bibinfo {type} {Tech. Rep.}\ (\bibinfo {address} {Amsterdam, The
  Netherlands, The Netherlands},\ \bibinfo {year} {1995})\BibitemShut {NoStop}%
\bibitem [{\citenamefont {van~der Walt}\ \emph {et~al.}(2014)\citenamefont
  {van~der Walt}, \citenamefont {{S}ch\"onberger}, \citenamefont
  {{Nunez-Iglesias}}, \citenamefont {{B}oulogne}, \citenamefont {{W}arner},
  \citenamefont {{Y}ager}, \citenamefont {{G}ouillart}, \citenamefont {{Y}u},\
  and\ \citenamefont {the scikit-image contributors}}]{vanderwalt2014}%
  \BibitemOpen
  \bibfield  {author} {\bibinfo {author} {\bibfnamefont {S.}~\bibnamefont
  {van~der Walt}}, \bibinfo {author} {\bibfnamefont {J.~L.}\ \bibnamefont
  {{S}ch\"onberger}}, \bibinfo {author} {\bibfnamefont {J.}~\bibnamefont
  {{Nunez-Iglesias}}}, \bibinfo {author} {\bibfnamefont {F.}~\bibnamefont
  {{B}oulogne}}, \bibinfo {author} {\bibfnamefont {J.~D.}\ \bibnamefont
  {{W}arner}}, \bibinfo {author} {\bibfnamefont {N.}~\bibnamefont {{Y}ager}},
  \bibinfo {author} {\bibfnamefont {E.}~\bibnamefont {{G}ouillart}}, \bibinfo
  {author} {\bibfnamefont {T.}~\bibnamefont {{Y}u}}, \ and\ \bibinfo {author}
  {\bibnamefont {the scikit-image contributors}},\ }\href {\doibase
  10.7717/peerj.453} {\bibfield  {journal} {\bibinfo  {journal} {PeerJ}\
  }\textbf {\bibinfo {volume} {2}},\ \bibinfo {pages} {e453} (\bibinfo {year}
  {2014})}\BibitemShut {NoStop}%
\bibitem [{\citenamefont {Bradski}(2000)}]{bradski2000}%
  \BibitemOpen
  \bibfield  {author} {\bibinfo {author} {\bibfnamefont {G.}~\bibnamefont
  {Bradski}},\ }\href@noop {} {\bibfield  {journal} {\bibinfo  {journal} {Dr.
  Dobb's Journal of Software Tools}\ } (\bibinfo {year} {2000})}\BibitemShut
  {NoStop}%
\bibitem [{\citenamefont {Canny}(1987)}]{canny1987}%
  \BibitemOpen
  \bibfield  {author} {\bibinfo {author} {\bibfnamefont {J.}~\bibnamefont
  {Canny}},\ }in\ \href@noop {} {\emph {\bibinfo {booktitle} {{Readings in
  Computer Vision}}}}\ (\bibinfo  {publisher} {Elsevier},\ \bibinfo {year}
  {1987})\ pp.\ \bibinfo {pages} {184--203}\BibitemShut {NoStop}%
\bibitem [{\citenamefont {Jaskula}\ \emph {et~al.}(2017)\citenamefont
  {Jaskula}, \citenamefont {Bauch}, \citenamefont {Arroyo-Camejo},
  \citenamefont {Lukin}, \citenamefont {Hell}, \citenamefont {Trifonov},\ and\
  \citenamefont {Walsworth}}]{JCsuperres}%
  \BibitemOpen
  \bibfield  {author} {\bibinfo {author} {\bibfnamefont {J.-C.}\ \bibnamefont
  {Jaskula}}, \bibinfo {author} {\bibfnamefont {E.}~\bibnamefont {Bauch}},
  \bibinfo {author} {\bibfnamefont {S.}~\bibnamefont {Arroyo-Camejo}}, \bibinfo
  {author} {\bibfnamefont {M.~D.}\ \bibnamefont {Lukin}}, \bibinfo {author}
  {\bibfnamefont {S.~W.}\ \bibnamefont {Hell}}, \bibinfo {author}
  {\bibfnamefont {A.~S.}\ \bibnamefont {Trifonov}}, \ and\ \bibinfo {author}
  {\bibfnamefont {R.~L.}\ \bibnamefont {Walsworth}},\ }\href {\doibase
  10.1364/OE.25.011048} {\bibfield  {journal} {\bibinfo  {journal} {Opt.
  Express}\ }\textbf {\bibinfo {volume} {25}},\ \bibinfo {pages} {11048}
  (\bibinfo {year} {2017})}\BibitemShut {NoStop}%
\bibitem [{\citenamefont {de~Groot}\ \emph {et~al.}(2018)\citenamefont
  {de~Groot}, \citenamefont {Fabian}, \citenamefont {Béguin}, \citenamefont
  {Reith}, \citenamefont {Barnhoorn},\ and\ \citenamefont
  {Hilgenkamp}}]{lennartSQUID}%
  \BibitemOpen
  \bibfield  {author} {\bibinfo {author} {\bibfnamefont {L.~V.}\ \bibnamefont
  {de~Groot}}, \bibinfo {author} {\bibfnamefont {K.}~\bibnamefont {Fabian}},
  \bibinfo {author} {\bibfnamefont {A.}~\bibnamefont {Béguin}}, \bibinfo
  {author} {\bibfnamefont {P.}~\bibnamefont {Reith}}, \bibinfo {author}
  {\bibfnamefont {A.}~\bibnamefont {Barnhoorn}}, \ and\ \bibinfo {author}
  {\bibfnamefont {H.}~\bibnamefont {Hilgenkamp}},\ }\href {\doibase
  10.1002/2017GL076634} {\bibfield  {journal} {\bibinfo  {journal} {Geophysical
  Research Letters}\ }\textbf {\bibinfo {volume} {45}},\ \bibinfo {pages}
  {2995} (\bibinfo {year} {2018})}\BibitemShut {NoStop}%
\bibitem [{\citenamefont {Rukhin}\ \emph {et~al.}(2010)\citenamefont {Rukhin},
  \citenamefont {Soto}, \citenamefont {Nechvatal}, \citenamefont {Smid},
  \citenamefont {Barker}, \citenamefont {Leigh}, \citenamefont {Levenson},
  \citenamefont {Vangel}, \citenamefont {Banks}, \citenamefont {Heckert},
  \citenamefont {Dray},\ and\ \citenamefont {Vo}}]{NISTrng}%
  \BibitemOpen
  \bibfield  {author} {\bibinfo {author} {\bibfnamefont {A.}~\bibnamefont
  {Rukhin}}, \bibinfo {author} {\bibfnamefont {J.}~\bibnamefont {Soto}},
  \bibinfo {author} {\bibfnamefont {J.}~\bibnamefont {Nechvatal}}, \bibinfo
  {author} {\bibfnamefont {M.}~\bibnamefont {Smid}}, \bibinfo {author}
  {\bibfnamefont {E.}~\bibnamefont {Barker}}, \bibinfo {author} {\bibfnamefont
  {S.}~\bibnamefont {Leigh}}, \bibinfo {author} {\bibfnamefont
  {M.}~\bibnamefont {Levenson}}, \bibinfo {author} {\bibfnamefont
  {M.}~\bibnamefont {Vangel}}, \bibinfo {author} {\bibfnamefont
  {D.}~\bibnamefont {Banks}}, \bibinfo {author} {\bibfnamefont
  {A.}~\bibnamefont {Heckert}}, \bibinfo {author} {\bibfnamefont
  {J.}~\bibnamefont {Dray}}, \ and\ \bibinfo {author} {\bibfnamefont
  {S.}~\bibnamefont {Vo}},\ }\href@noop {} {\bibfield  {journal} {\bibinfo
  {journal} {NIST Special Publication 800-22 Revision 1a}\ } (\bibinfo {year}
  {2010})}\BibitemShut {NoStop}%
\bibitem [{\citenamefont {Tauxe}(2018)}]{tauxePaleomag}%
  \BibitemOpen
  \bibfield  {author} {\bibinfo {author} {\bibfnamefont {L.}~\bibnamefont
  {Tauxe}},\ }\href@noop {} {\emph {\bibinfo {title} {Essentials of
  {P}aleomagnetism}}}\ (\bibinfo  {publisher} {University of California
  Press},\ \bibinfo {year} {2018})\BibitemShut {NoStop}%
\bibitem [{\citenamefont {Harrison}\ and\ \citenamefont
  {Feinberg}(2009)}]{elements209}%
  \BibitemOpen
  \bibfield  {author} {\bibinfo {author} {\bibfnamefont {R.~J.}\ \bibnamefont
  {Harrison}}\ and\ \bibinfo {author} {\bibfnamefont {J.~M.}\ \bibnamefont
  {Feinberg}},\ }\href {\doibase 10.2113/gselements.5.4.209} {\bibfield
  {journal} {\bibinfo  {journal} {Elements}\ }\textbf {\bibinfo {volume} {5}},\
  \bibinfo {pages} {209} (\bibinfo {year} {2009})}\BibitemShut {NoStop}%
\bibitem [{\citenamefont {Gold}\ and\ \citenamefont
  {Tucker}(1989)}]{magneticStripePatent1}%
  \BibitemOpen
  \bibfield  {author} {\bibinfo {author} {\bibfnamefont {D.~G.}\ \bibnamefont
  {Gold}}\ and\ \bibinfo {author} {\bibfnamefont {F.~D.}\ \bibnamefont
  {Tucker}},\ }\href@noop {} {\enquote {\bibinfo {title} {Magnetic
  characteristic identification system},}\ } (\bibinfo {year} {1989}),\
  \bibinfo {note} {{US Patent 4,806,740}}\BibitemShut {NoStop}%
\bibitem [{\citenamefont {Indeck}\ and\ \citenamefont
  {Muller}(1994)}]{magneticStripePatent2}%
  \BibitemOpen
  \bibfield  {author} {\bibinfo {author} {\bibfnamefont {R.~S.}\ \bibnamefont
  {Indeck}}\ and\ \bibinfo {author} {\bibfnamefont {M.~W.}\ \bibnamefont
  {Muller}},\ }\href@noop {} {\enquote {\bibinfo {title} {Method and apparatus
  for fingerprinting magnetic media},}\ } (\bibinfo {year} {1994}),\ \bibinfo
  {note} {{US Patent 5,365,586}}\BibitemShut {NoStop}%
\bibitem [{\citenamefont {Samid}\ and\ \citenamefont
  {Wnek}(2019)}]{rockOfRandomness}%
  \BibitemOpen
  \bibfield  {author} {\bibinfo {author} {\bibfnamefont {G.}~\bibnamefont
  {Samid}}\ and\ \bibinfo {author} {\bibfnamefont {G.~E.}\ \bibnamefont
  {Wnek}},\ }\href {\doibase 10.1557/mrc.2019.8} {\bibfield  {journal}
  {\bibinfo  {journal} {MRS Communications}\ }\textbf {\bibinfo {volume} {9}},\
  \bibinfo {pages} {67–76} (\bibinfo {year} {2019})}\BibitemShut {NoStop}%
\end{thebibliography}

%

\clearpage


\section*{Supplementary material}

# Supplemental material for "A physically unclonable function using NV diamond magnetometry and micromagnet arrays"

Pauli Kehayias,[1, *] Ezra Bussmann,[1, 2] Tzu-Ming Lu,[1, 2] and Andrew M. Mounce[1, 2]

[1]*Sandia National Laboratories, Albuquerque, New Mexico 87185, USA*
[2]*Center for Integrated Nanotechnologies, Sandia National Laboratories, Albuquerque, New Mexico 87123, USA*

## EXAMPLE MAGNETIC DIPOLE MAPS MEASURED ALONG THE [111] AND $z$ DIRECTIONS

In our experiment we used projection magnetic microscopy (PMM) [S1] to measure the magnetic field projection along the NV axis (the [111] crystallographic direction). For our diamond chip, the unit vector for the [111] direction is $\{0, \sqrt{\frac{2}{3}}, \sqrt{\frac{1}{3}}\}$, and points $\sim 35°$ out-of-plane. For a magnetic dipole located at $\{0, 0, -z_0\}$ with a magnetic moment $\vec{m} = \{m_x, m_y, m_z\}$, the magnetic field (evaluated in the $x$-$y$ plane) is

$$B_x(x,y) = \frac{\mu_0}{4\pi} \frac{3x(m_y y + m_z z_0) + m_x(2x^2 - y^2 - z_0^2)}{(x^2 + y^2 + z_0^2)^{\frac{5}{2}}}, \tag{S1}$$

$$B_y(x,y) = \frac{\mu_0}{4\pi} \frac{3y(m_x x + m_z z_0) - m_y(x^2 - 2y^2 + z_0^2)}{(x^2 + y^2 + z_0^2)^{\frac{5}{2}}}, \tag{S2}$$

$$B_z(x,y) = \frac{\mu_0}{4\pi} \frac{3z_0(m_x x + m_y y) - m_z(x^2 + y^2 - 2z_0^2)}{(x^2 + y^2 + z_0^2)^{\frac{5}{2}}}, \tag{S3}$$

where $\mu_0 = 4\pi \times 10^{-7}$ T·m/A. Figure S1 shows calculated $B_z(x,y)$ and $B_{111}(x,y) = \sqrt{\frac{2}{3}} B_y(x,y) + \sqrt{\frac{1}{3}} B_z(x,y)$ magnetic maps for a dipole along the $y$-axis. As expected, a magnetic dipole model describes the measured micromagnet field maps shown in the main text. Since a $B_z(x,y)$ map is simpler to analyze (with two equal-sized lobes) and more compact than a $B_{111}(x,y)$ map, we calculate $B_z(x,y)$ using the expressions below. This simplifies our magnetic-to-bit string conversion and improves the micromagnet spatial separation, allowing us to pack the micromagnets closer together without overlap.

## CALCULATING A $B_z$ MAP FROM A $B_{111}$ MAP

In Ref. [S2], the authors show how to generate the $B_x(x,y)$ and $B_y(x,y)$ magnetic components from a measured $B_z(x,y)$ map. After calculating the 2D Fourier transform $b_z(k_x, k_y)$, where $k_x$ and $k_y$ are spatial frequencies, they write

$$b_x(k_x, k_y) = \frac{-ik_x b_z(k_x, k_y)}{\sqrt{k_x^2 + k_y^2}}, \quad b_y(k_x, k_y) = \frac{-ik_y b_z(k_x, k_y)}{\sqrt{k_x^2 + k_y^2}}. \tag{S4}$$

In our experiment we measure $B_{111}(x,y) = \sqrt{\frac{2}{3}} B_y(x,y) + \sqrt{\frac{1}{3}} B_z(x,y)$, and the Fourier transform of both sides yields $b_{111}(k_x, k_y) = \sqrt{\frac{2}{3}} b_y(k_x, k_y) + \sqrt{\frac{1}{3}} b_z(k_x, k_y)$. Solving the above expressions, we get

$$b_z(k_x, k_y) = \frac{b_{111}(k_x, k_y)}{\sqrt{\frac{2}{3}} \frac{-ik_y}{\sqrt{k_x^2 + k_y^2}} + \sqrt{\frac{1}{3}}}. \tag{S5}$$

After taking the inverse Fourier transform, we obtain the desired $B_z(x,y)$ map. Figure S2 shows an example $B_z$ map of an isolated micromagnet calculated from a measured $B_{111}$ map using this technique.

## MICROMAGNET PUF IMAGING WITH A THINNER NV LAYER

As mentioned in the main text, the NV layer thickness is one thing we can adjust to optimize the micromagnet readout performance. A thicker NV layer means brighter NV fluorescence and a better magnetic noise floor, but

this comes at the cost of larger standoff distance, worse spatial resolution, and weaker micromagnet field strength. Since the magnetic field from a dipole increases with distance cubed, a thinner NV layer can yield an improved signal-to-noise ratio if the field strength improvement is larger than the sensitivity reduction. However, this only works up to a point, since ensuring that a few-millimeter diamond chip lies on top of the micromagnet array with no air gap is difficult.

Here we compare two diamond samples: Sample A (which has a 4 µm NV layer) and Sample B (which has a 0.15 µm NV layer). Table S1 lists the relevant properties for the two diamonds used in this comparison. For Sample A (which we used for the measurements in the main text) we found a 3.6 µm standoff distance. Since this standoff distance is comparable to the NV layer thickness, it is not obvious how much the NV layer thickness contributes to the standoff distance compared to the air gap thickness. To overcome this ambiguity, we assess the standoff distance using a second diamond chip Sample B, for which the NV layer thickness contribution is negligible. Note that although the two diamonds differ in several aspects (such as NV layer formation method, $^{13}$C density, nitrogen isotope, and magnetic sensitivity), these differences should not affect the conclusion.

Figure S3 shows the standoff distance assessment with sparsely-placed $0.25 \times 1$ µm$^2$ micromagnets, for which we find a 1.9 µm mean standoff distance. This standoff distance is larger than the micromagnet dimensions and the NV layer thickness, providing a measurement of the air gap only. This suggests that reducing the NV layer thickness below ∼2 µm is impractical, since in practice the typical NV-micromagnet separation will not be shorter than ∼2 µm due to the air gap.

Note that the optical diffraction limit in our microscope is quite coarse (1.4 µm), but after simulating how optical diffraction affects the measured magnetic images for dipoles at different standoff heights, we determined that optical diffraction only matters if the standoff distance is less than ∼0.5 µm.

## BIT STRING RANDOMNESS STATISTICAL TEST RESULTS

Table S2 shows the full results of a suite of fifteen statistical tests of bit string randomness published by NIST [S3]. We applied these statistical tests for the binary string extracted from Fig. 2a in the main text. This table includes the calculated $p$-values and additional comments for certain statistical tests. Several of the statistical tests require a longer binary string than we measured ($10^4$ bits), and thus produced no output. Note that the Cumulative Sums Test, Serial Test, and Non-Overlapping Template Test each provide multiple $p$-value outputs.

---

* pmkehay@sandia.gov
[S1] D. R. Glenn, R. R. Fu, P. Kehayias, D. Le Sage, E. A. Lima, B. P. Weiss, and R. L. Walsworth, Geochemistry, Geophysics, Geosystems **18**, 3254 (2017).
[S2] E. A. Lima and B. P. Weiss, Journal of Geophysical Research: Solid Earth **114** (2009).
[S3] A. Rukhin, J. Soto, J. Nechvatal, M. Smid, E. Barker, S. Leigh, M. Levenson, M. Vangel, D. Banks, A. Heckert, J. Dray, and S. Vo, NIST Special Publication 800-22 Revision 1a (2010).

| Sample | Dimensions | NV layer thickness | [N] in layer | [$^{13}$C] in layer | Noise floor in $1 \times 1$ µm$^2$ area | Standoff |
|---|---|---|---|---|---|---|
| A | $4 \times 4 \times 0.5$ mm$^3$ | 4 µm | 20 ppm ($^{14}$N grown in) | 0.001% | 7 µT after 1 s | 3.6 µm |
| B | $2 \times 2 \times 0.5$ mm$^3$ | 0.15 µm | 45 ppm ($^{15}$N implant) | 1.1% | ∼70 µT after 1 s | 1.9 µm |

**Supplementary Table** S1. Relevant properties for Sample A (used in the main text) and Sample B (which has a thinner 0.15 µm NV layer).

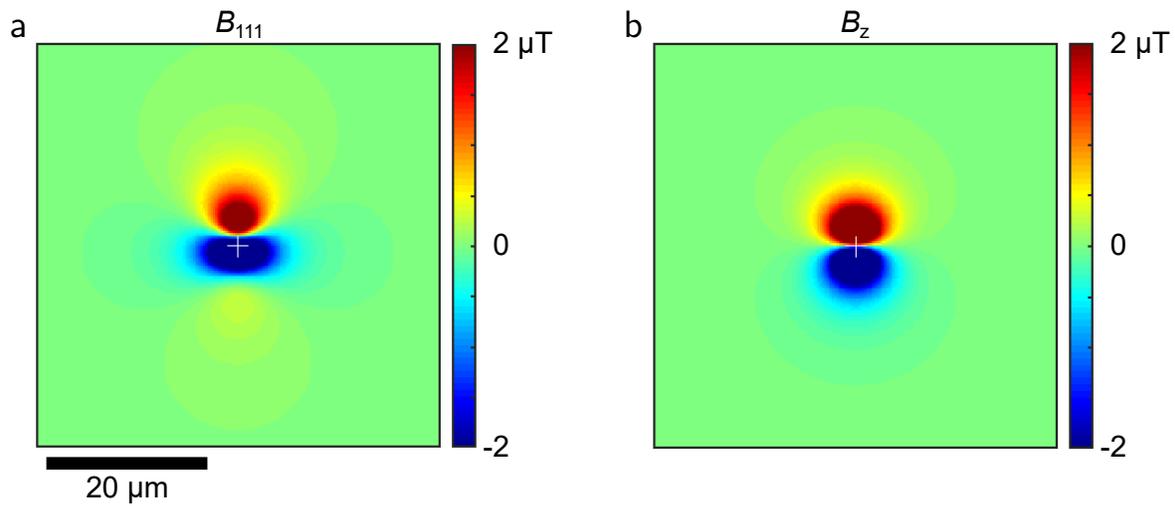

**Supplementary Figure** S1. (a) Simulated $B_{111}$ map for a magnetic dipole along the $y$ direction ($3\times10^{-15}$ J/T moment, 3.5 µm depth). (b) Simulated $B_z$ map for the same magnetic dipole. Compared to the $B_{111}$ map, the $B_z$ map is simpler (it has symmetric positive and negative regions), has a more obvious dipole $\{x, y\}$ location, and is more compact.

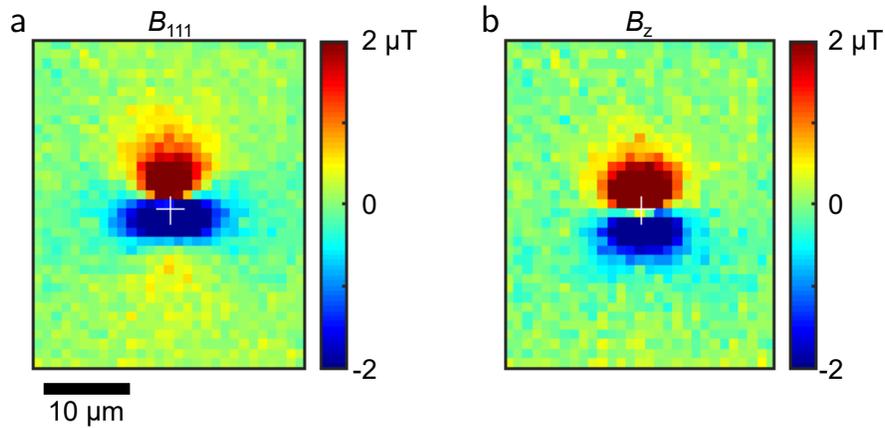

**Supplementary Figure** S2. (a) The $B_{111}$ magnetic field map of a 0.25×1 µm² isolated micromagnet (measured with a 4 µm NV layer), with $3.6\times10^{-15}$ J/T magnetic moment along $y$. (b) The calculated $B_z$ map of the same micromagnet calculated from (a) using Eq. S5.

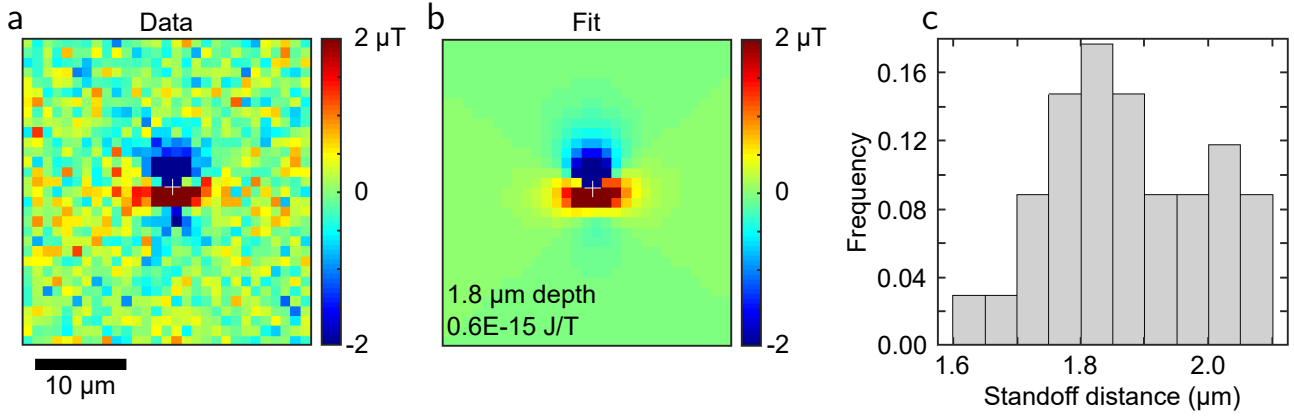

**Supplementary Figure** S3. (a) Example $B_{111}$ data of a 0.25×1 µm² micromagnet measured with a 0.15 µm NV layer. (b) Magnetic dipole fit for the data in (a). (c) Histogram of 34 fitted dipole standoff heights, showing a mean height of 1.9 µm with a 0.1 µm standard deviation.

| Statistical Test | $p$-value | Result | Comments |
|---|---|---|---|
| Frequency | 0.023821 | Pass | |
| Block Frequency | 0.036802 | Pass | |
| Runs | 0.000000 | Fail | Bit string values fluctuate too slowly |
| Longest Run | 0.465732 | Pass | |
| Rank | 0.862457 | Pass | |
| Discrete Fourier Transform | 0.066457 | Pass | |
| Non-Overlapping Template | - | - | 138 out of 148 tests pass |
| Overlapping Template | 0.241223 | Pass | |
| Maurer's Universal | - | - | Too few bits to evaluate |
| Linear Complexity | 0.543677 | Pass | |
| Serial ($\nabla \psi_m^2$) | 0.059629 | Pass | |
| Serial ($\nabla^2 \psi_m^2$) | 0.311516 | Pass | |
| Approximate Entropy | 0.000503 | Fail | |
| Cumulative Sums Forward | 0.027787 | Pass | |
| Cumulative Sums Reverse | 0.036550 | Pass | |
| Random Excursions | - | - | Too few bits to evaluate |
| Random Excursions Variant | - | - | Too few bits to evaluate |

**Supplementary Table** S2. A table summarizing the statistical test results for bit string randomness for the micromagnet array shown in Fig. 2a in the main text [S3].



\end{document}